\documentclass[seceq]{ptptex}
\usepackage{epsfig}

\title{
Quantum Hall Systems Studied by the Density Matrix \\ 
Renormalization Group Method}

\author{
Naokazu \textsc{Shibata}%
}

\inst{
Depertment of Physicas, Tohoku University, Aoba, Aoba-ku, Sendai,
Miyagi 980-8578}

\abst{
The ground-state and low-energy excitations of quantum Hall systems
are studied by the density matrix renormalization group (DMRG) method.
From the ground-state pair correlation functions and low-energy 
excitions, the ground-state phase diagram is determined, which
consists of incompressible liquid states, 
Fermi liquid type compressible liquid states, and many kinds of CDW 
states called stripe, bubble and Wigner crystal. 
The spin transition and the domain formation 
are studied at $\nu=2/3$.
The evolution from composite fermion liquid state to an 
excitonic state in bilayer systems is investigated at 
total filling factor $\nu=1$.
}

\begin{document}

\maketitle

\section{Introduction}
In two dimensional systems, applying perpendicular magnetic
field strongly modifies the wave function of electrons
leading to many interesting phenomena at low temperatures. 
The fractional quantum Hall effects\cite{Laughlin}  are typical example, 
where incompressible ground states are realized  
only at some fractional fillings of Landau levels.\cite{Tsui,FQHEex}
Since fractional quantum Hall effects are observed only 
in high quality samples, the Coulomb interaction between 
the electrons is thought to be essential rather than 
random potentials from impurities.
This is contrasted with the case of integer quantum Hall effect
where random potentials are essential.
The importance of the Coulomb interaction in high magnetic field  
is followed from
 the increase in the energy scale of the Coulomb interaction. 
The wave function is scaled by the magnetic length 
$\ell = \sqrt{\hbar/eB}$, 
which is equivalent to the classical cyclotron 
radius $r_c$ in the lowest Landau level.
The increase in the magnetic field decreases
the magnetic length and enhances the energy scale of the  
Coulomb interaction between the electrons, $e^2/\varepsilon \ell$. 

At typical magnetic field of 10T, $\ell$ is about 8nm, 
which is still much larger than the atomic length of 0.1nm. 
Since the conduction electrons are on the positive 
background charge from ions over length scale of $\ell$, 
the positive charge may be simplified to be uniform. 
Then the system is equivalent to the electron gas 
in a magnetic field, and $\ell$ becomes unique length
scale of the system.

In the magnetic field, the kinetic energy is scaled by the 
cyclotron frequency $\omega_c=eB/m$, which is 
determined by the magnetic field $B$. 
The quantization of the wave function 
discretizes the classical cyclotron radius $r_c$, 
which also discretizes the kinetic energy 
and makes Landau levels $E=\hbar \omega_c (n+1/2)$.
This means that the macroscopic number of electrons have
the same energy in each Landau level, and
large-scale degeneracy appears in the ground state.
This macroscopic degeneracy is lifted by
the Coulomb interaction between the electrons and 
various types of liquid states\cite{RezHal,Grei,Moor,Pan,Eis} 
and CDW states\cite{Kou1,Lill,Du,Coop} are realized 
depending on the filling of the Landau levels.

Since the ground state has macroscopic 
degeneracy in the limit of weak Coulomb interaction,
standard perturbation theories are not useful. 
Thus numerical diagonalizations of the many body 
Hamiltonian have been used to study this system.
Since numerical representation of the Hamiltonian
needs complete set of many body basis states,
we divide the system into unit cells with 
finite number of electrons in each cell.
The properties of the infinite system are obtained by
the finite size scalings. 
However, the number of many body basis states 
increases exponentially with the number of electrons.
For example, when we study the ground state at $\nu=1/3$ with 
18 electrons, each unit cell has 54 degenerated orbitals. 
The number of many body basis states is given by the combination of
occupied and unoccupied orbitals, $_{54}C_{18} \sim 10^{14}$,
which is practically impossible to manage by using standard 
numerical method such as exact diagonalization.

To study systems with typically more than 10 electrons, 
we need to reduce the number of many body basis states. 
For this purpose, we use the density matrix renormalization 
group (DMRG) method, which was originally developed by  
S. White in 1992.\cite{White1,White2}
This method is a kind of variational method combined with 
a real space renormalization group method, which enables 
us to obtain the ground-state wave function of large-size 
systems with controlled 
high accuracy within a restricted number of many body basis states. 
The DMRG method has excellent features
compared with other standard numerical methods.
In contrast to the quantum Monte Carlo method, 
the DMRG method is free from statistical errors and 
the negative sign problem, which inhibit convergence 
of physical quantities at low temperatures.
Compared with the exact diagonalization method, the DMRG method 
has no limitation in the size of system.
The error in the DMRG calculation  
comes from restrictions of the number of basis states,
which is systematically controlled by the density 
matrix calculated from the ground-state wave function,
and the obtained results are easily improved by 
increasing the number of basis states retained in the system.

The application of the DMRG method to two-dimensional quantum 
systems is a challenging subject and many algorithms have been 
proposed. Most of them use mappings on to effective
one-dimensional models with long-range interactions.
However, the mapping from two-dimensional systems to 
one-dimensional effective models is not unique and proper 
mapping is necessary to keep high accuracy.
In two-dimensional systems under a perpendicular magnetic field,
all the one-particle wave functions  $\Psi_{N X}(x,y)$ 
are identified by the Landau level index $N$ and the 
x-component of the guiding center, $X$, in Landau gage. 
The guiding center is essentially the center coordinate of the 
cyclotron motion of the electron and it is natural to use $X$ as a
one-dimensional index of the effective model.
More importantly, $X$ is discretized in finite unit cell 
of $L_x\times L_y$ through the relation to $y$-momentum, 
$X=k_y \ell^2 $, 
which is discretized under the periodic boundary condition, 
$k_y=2\pi n/L_y$ with $n$ being an integer.
Therefore, the two-dimensional continuous systems in magnetic field
are naturally mapped on to effective one-dimensional 
lattice models, and we can apply the standard DMRG method.\cite{Shibata1}

This method was first applied to interacting 
electron systems in a high Landau level and 
the ground-state phase diagram, which consists of various
CDW states called stripe, bubble and Wigner crystal,
has been determined.\cite{Shibata2,Yoshioka} 
The ground state and low energy excitations in the lowest 
and the second lowest Landau levels have also been 
studied by the DMRG and the existence of 
various quantum liquid states such as Laughlin state and charge 
ordered states called Wigner crystal have been confirmed
and new stripe state has been proposed.\cite{Shibata3,Shibata6}

In the following, we first explain the effective one-dimensional 
Hamiltonian used in the above studies and then show the results 
obtained for the spin polarized single layer system.
We next review recent study on the spin transition and domain 
formation at $\nu=2/3$,\cite{Shibata4} and 
finally explain the results on bilayer quantum Hall systems 
at $\nu=1$,\cite{Shibata5}
where crossover from a Fermi liquid state to an excitonic 
incompressible state occurs.

\section{DMRG method}
Here we briefly describe how the effective 1D Hamiltonian is
obtained from 2D quantum Hall systems.\cite{Shibata1}
To describe the many body Hamiltonian for a interacting system,  
we first need to define one-particle basis states.
Here, we use the eigenstates of free electrons 
in a magnetic field as one-particle basis states and 
represent the wave function $\Psi_{N X}(x,y)$ in Landau gauge:
\begin{equation}
\label{BWF}
\Psi_{N X}(x,y) = C_{N} \exp{\left[i {k_y y} -\frac{(x-X)^2}
{2\ell^2}\right]} H_N\left[\frac{x-X}{\ell}\right],
\end{equation}
where $H_N$ are Hermite
polynomials and $C_{N}$ is the normalization constant.
Then all the eigenstates $\Psi_{N X}(x,y)$ are specified 
using two independent parameters $N$ and $X$;
$N$ is the Landau level index and $X$ is the $x$-component
of the guiding center coordinates of the electron.
Since the guiding center $X$ is related to the momentum 
$k_y$ as  $X=k_y\ell^2$, and $k_y$ is discretized under 
the periodic boundary conditions, 
the guiding center $X$ takes only discrete values
\begin{equation}
X_n=2\pi\ell^2 n/ L_y,
\end{equation}
where $L_y$ is the length of the unit cell in the $y$-direction.

If we fix the Landau level index $N$, 
all the one-particle states are 
specified by one-dimensional discrete parameter $X_n$.
Since many body basis states are product states of 
one-particle states, they are also described by the 
combinations of $X_n$ of electrons in the system. 
Thus the system can be mapped on to an effective 
one-dimensional lattice model.

The macroscopic degeneracy in the ground state of 
free electrons in partially filled Landau level
is lifted by the Coulomb interaction
\begin{equation}
V(r)= \frac{e^2}{\epsilon r}.
\end{equation}
The Coulomb interaction makes correlations between 
the electrons and stabilizes various types of ground states
depending on the filling $\nu$ of Landau levels. 
When the magnetic field is strong enough so that 
the Landau level splitting is sufficiently
large compared with the typical Coulomb interaction
$e^2/(\epsilon \ell)$,
the electrons in fully occupied Landau 
levels are inert and the ground state is
determined only by the electrons in the top most
partially filled Landau level.

The Hamiltonian is then written by
\begin{equation}
\label{2DH}
H= S \sum_n c_n^\dagger c_n + 
\frac{1}{2}\sum_{n_1} \sum_{n_2} \sum_{n_3} \sum_{n_4} 
A_{n_1 n_2 n_3 n_4} c_{n_1}^\dagger  c_{n_2}^\dagger c_{n_3} c_{n_4},
\end{equation}
where we have imposed periodic boundary conditions in both
$x$- and $y$-directions, and
$S$ is the classical Coulomb energy of Wigner crystal
with a rectangular unit cell of $L_x \times L_y$\cite{QHHS}.
$c_n^\dagger$ is the creation operator of the electron represented 
by the wave function defined in equation (\ref{BWF}) with $X=X_n$. 
$A_{n_1 n_2 n_3 n_4}$ are the matrix elements of the Coulomb 
interaction defined by
\begin{eqnarray}
A_{n_1 n_2 n_3 n_4}&=&\delta'_{n_1+n_2,n_3+n_4}\frac{1}{L_xL_y}
\sum_{\mib q} \delta'_{n_1-n_4,q_yL_y/2\pi}\frac{2\pi e^2}{\epsilon q}
 \nonumber\\
&& {\mbox{\hspace{0.5cm}}}\times\left[L_N(q^2\ell^2/2)\right]^2
\exp{ \left[-\frac{q^2 \ell^2}{2}-i(n_1-n_3)\frac{q_xL_x}{M} \right] } ,
\end{eqnarray}
where $L_N(x)$ are Laguerre polynomials with $N$ being the Landau level
index\cite{Yoshioka}. 
$\delta_{n_1,n_2}' = 1$ when $n_1=n_2 (\mbox{mod}\ M)$ with
$M$ being the number of one-particle states in the unit cell, 
which is given by the area of the unit cell
$2\pi M \ell^2=L_xL_y$.

\begin{figure}[t]
\begin{center}
\epsfxsize=110mm \epsffile{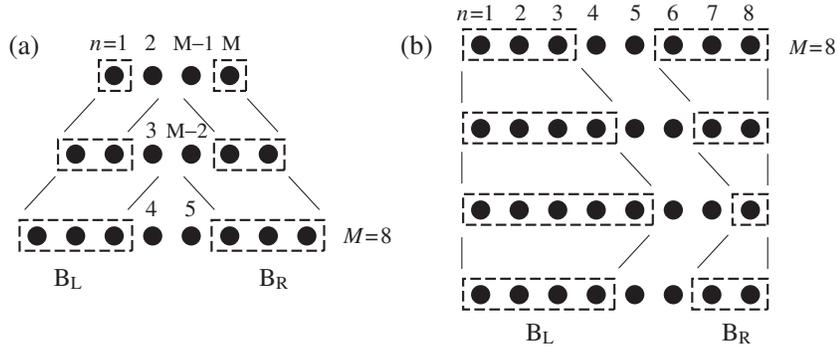}
\caption{\label{DMRG}
Schematic diagrams for (a) infinite system algorithm 
and (b) finite system algorithm of the DMRG method.
 $\bullet$ represents a one-particle orbital 
in a given Landau level. B$_L$ and B$_R$ are left and right blocks, 
respectively.}
\end{center}
\end{figure}

In order to obtain the ground-state wave function
we apply the DMRG method.\cite{Shibata1}
As shown in Fig.~\ref{DMRG} (a), we start from a small-size 
system consisting of
only four one-particle orbitals whose indices $n$
are 1, 2, $M-1$, and $M$, and we
calculate the ground-state wave function. We then construct
the left block containing one-particle orbitals of $n=1$ and 2, and the 
right block containing  $n=M-1$ and $M$ by using eigenvectors
of the density matrices which are calculated from the 
ground-state wave function. 
We then add two one-particle orbitals $n=3$ and $M-2$ between the
two blocks and repeat the above procedure until
$M$ one-particle orbitals are included in the system.
We then apply the finite system algorithm of the DMRG
shown in Fig.~\ref{DMRG} (b)
to refine the ground-state wave function.
After we have obtained the convergence, 
we calculate correlation functions to identify the ground state.

The ground-state pair correlation function $g({\mib r})$ in 
guiding center coordinates is defined by 
\begin{equation}
g({\mib r})=\frac{L_xL_y}{N_e(N_e-1)}
\langle \Psi | \sum_{i\ne j} \delta({\mib r-\mib R}_i+{\mib R}_j) | \Psi \rangle ,
\end{equation}
where ${\mib R}_i$ is the guiding center coordinate of the $i$th
electron, and it is calculated from the following equation
\begin{eqnarray}
g({\mib r})&=& 
\frac{1}{N_e(N_e-1)}\sum_{\mib q}\sum_{n_1,n_2,n_3,n_4} \exp
\left[ i{\mib q \cdot\mib r}-\frac{q^2\ell^2}{2}-i(n_1-n_3)\frac{q_xL_x}{M}
\right] \times \nonumber\\
&& \mbox{\hspace{3.5cm}} \delta'_{n_1-n_4,q_yL_y/2\pi} \langle 
\Psi |  c_{n_1}^\dagger  c_{n_2}^\dagger c_{n_3} c_{n_4} | \Psi \rangle,
\end{eqnarray}
where $\Psi$ is the ground state and $N_e$ is the total number of 
electrons.

\begin{figure}[t]
\begin{center}
\epsfxsize=75mm \epsffile{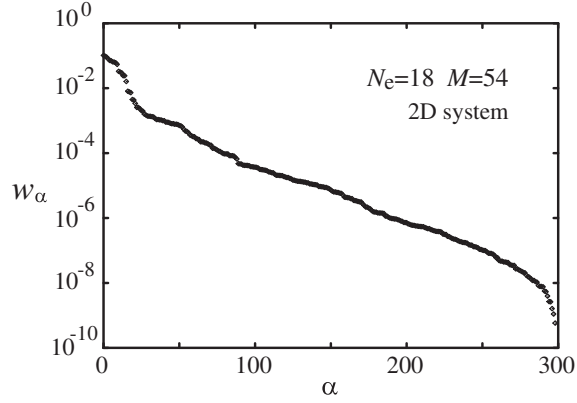}
\caption{\label{Fig_2DDM}
Eigenvalues $w_\alpha$ of the density matrix for 
two-dimensional system of 54 orbitals with 18 electrons.
Sum of $w_\alpha$ is equivalent to
the norm of the ground-state wave function 
and normalized to be unity.}
\end{center}
\end{figure}

The accuracy of the results depends on the 
distribution of eigenvalues of the density matrix.
A typical example of the eigenvalues of the 
density matrix for system of $M=54$ with $18$ electrons
is shown in Fig.~\ref{Fig_2DDM},
which shows an exponential decrease of eigenvalues $w_\alpha$.
In this case accuracy of $10^{-4}$ is obtained
by keeping more than one hundred states in each block. 

\section{Single layer system}

\begin{figure}[t]
\begin{center}
\epsfxsize=75mm \epsffile{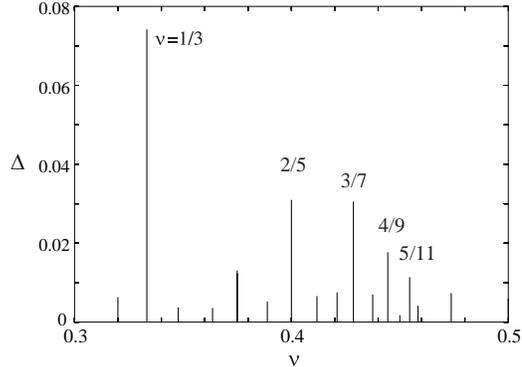}
\caption{\label{Fig_Gap}
The lowest excitation gap at various $\nu$ in 
the lowest Landau level. 
Relatively large excitation gap is obtained at
fractional fillings $\nu=n/(2n+1)$.
The excitation gap is in units of $e^2/(\epsilon \ell)$.
}
\end{center}
\end{figure}

Here we present diverse ground states obtained by the  
DMRG method applied to the single layer quantum Hall systems.
In the limit of strong magnetic field, the electrons 
occupy only the lowest Landau level $N=0$. In this limit,
fractional quantum Hall effect (FQHE) has been 
observed at various fractional fillings\cite{FQHEex}.
The FQHE state is characterized by incompressible liquid 
with a finite excitation gap\cite{Laughlin}.

\begin{figure}
\begin{center}
\epsfxsize=75mm \epsffile{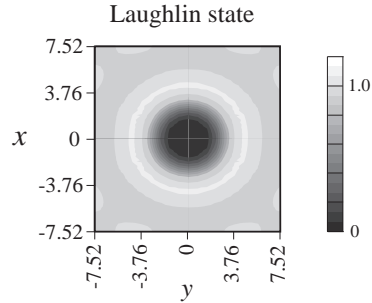}
\caption{\label{Fig_Lau}
Pair correlation function $g({\bf r})$ at $\nu=1/3$ in the lowest 
Landau level. The length is in units of $\ell$.
}
\end{center}
\end{figure}

These FQHE states are confirmed by the DMRG calculations,
where relatively large excitation gaps are obtained at
various fillings between $\nu=1/2$ and 3/10\cite{Shibata3}
as shown in Fig.~\ref{Fig_Gap}.
We clearly find large excitation gaps at fractional 
fillings $\nu=1/3,2/5,3/7,4/9$ and $5/11$,
which correspond to primary series of the FQHE at 
$\nu=n/(2n+1)$. 
The pair correlation function at $\nu=1/3$ is 
presented in Fig.~\ref{Fig_Lau}, which shows a
circularly symmetric correlation consistent with the 
Laughlin's wave function.\cite{Laughlin}

\begin{figure}
\begin{center}
\epsfxsize=95mm \epsffile{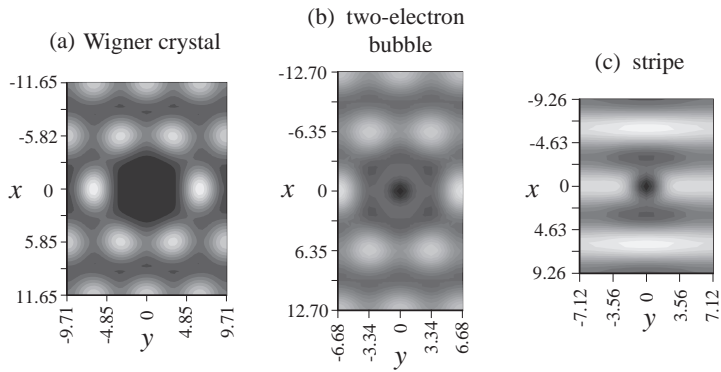}
\caption{\label{Fig_CDW}
Pair correlation functions $g({\bf r})$ in guiding center coordinates.
(a) Wigner crystal realized in an excited state at $\nu=1/6$ in 
the lowest Landau level. The number of electrons in the unit cell $N_e$ is 12.
(b) Two-electron bubble state at $\nu=8/27$ in $N=2$ Landau level. 
$N_e=16$.
(c) Stripe state at $\nu=3/7$ in $N=2$ Landau level. $N_e=18$.
}
\end{center}
\end{figure}

In the limit of low filling $\nu\rightarrow 0$, mean 
separation between the electrons becomes much longer than the 
typical length-scale of the one-particle wave function.
In this limit the quantum fluctuations are not important 
and electrons behave as classical point charges.
The ground state is then expected to be the Wigner crystal. 
The formation of the Wigner crystal is also confirmed by 
the DMRG calculations at low fillings
as shown in Fig.~\ref{Fig_CDW} (a).
The $\nu$-dependence of the low energy spectrum shows
that the first-order transition to Wigner crystal occurs
at $\nu\sim 1/7$.\cite{Shibata3}

With decreasing magnetic field, electrons occupy higher
Landau levels.
In high Landau levels, the one-particle wave function
extends over space leading to effective long range
exchange interactions between the electrons. 
The long range interaction 
stabilizes CDW ground states and 
various types of CDW states called stripe and bubble are 
predicted by Hartree-Fock theory.\cite{Kou1}
These CDW states are confirmed by the DMRG calculations 
as shown in Figs.~\ref{Fig_CDW} (b) and (c),
where two-electron bubble state and stripe state are
obtained at $\nu=8/27$ and $3/7$, respectively,  
in the $N=2$ Landau level.
Although the CDW structures are similar to those
obtained in the Hartree-Fock calculations, the ground 
state energy and 
the phase diagram are significantly different\cite{Shibata2}.
The DMRG results are consistent with recent 
experiments\cite{Lill},
and the discrepancy is due to the quantum fluctuations 
neglected in the Hartree-Fock calculations.

\begin{figure}
\begin{center}
\epsfxsize=75mm \epsffile{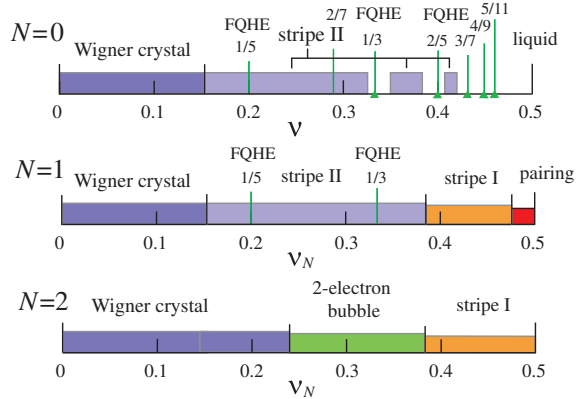}
\caption{\label{Phase}
The ground state phase diagram obtained by the DMRG method.
$N$ is the Landau level index and $\nu_N$ in the filling
factor of the $N$th Landau level.
}
\end{center}
\end{figure}

\begin{figure}
\begin{center}
\epsfxsize=120mm \epsffile{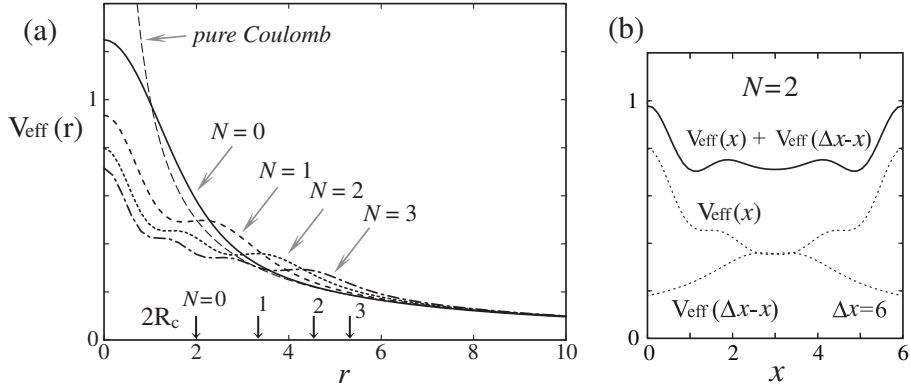}
\caption{\label{effective}
(a) Effective interaction between the electrons in the
$N$th Landau level. $R_c$ is the classical cyclotron radius.
(b) Coulomb potential made by two electrons separated by $\Delta x$.
}
\end{center}
\end{figure}

The ground-state phase diagram obtained by the DMRG
is shown in Fig.~\ref{Phase}. 
In the lowest Landau level, 
we find many liquid states at fractional 
fillings and around $\nu=1/2$. 
Nevertheless, 
CDW states dominate over the whole range of filling 
in higher Landau levels.
This difference in the ground state phase diagram 
comes from different effective interactions between the electrons.
In the lowest Landau level, the one particle 
wave function is localized within the magnetic length
$\ell$, that yields strong short-range repulsion between the
electrons. Since quantum liquid states such as Laughlin 
state are stabilized by the strong short-range repulsion,
liquid states are realized in the lowest Landau level.
In higher Landau levels, however, the wave function extends over
space with the increase in the classical cyclotron radius $r_c$.
Thus the short-range repulsion is reduced and 
liquid states become unstable.
As shown in Fig.~\ref{effective} (a),
the real space effective interaction between the electrons
in higher Landau levels
has a shoulder structure around the distance twice the classical 
cyclotron radius.
This structure of effective interaction makes 
minimum in the Coulomb potential near the guiding
center of the electron as shown in Fig.~\ref{effective} (b)
and stabilizes the clustering of electrons.
This is the reason why stripe and bubble states are
realized in higher Landau levels.\cite{Shibata6}

\section{Spin transitions}

In two dimensional systems, strong perpendicular magnetic field
completely quenches the kinetic energy of electrons.
Since the kinetic energy is independent of the spin polarization,
the exchange Coulomb interaction easily aligns the electron spin.
The ferromagnetic ground state at 
$\nu=1/q$ ($q$ odd) is thus realized even in the absence of the 
Zeeman splitting\cite{qhe-review}.
At the filling $\nu=2/3$ and $2/5$, however, the
paramagnetic ground states compete with the
ferromagnetic state, and the Zeeman splitting $\Delta_z=g\mu_BB$
induces a spin transition\cite{chacraborty}.
Such a spin transition in fractional quantum 
Hall states has been naively explained by the 
composite fermion theory.\cite{jain} 
The composite fermions are electrons coupled with even number
of fluxes.  These fluxes effectively reduces external
magnetic field and the $\nu=p/(2p\pm 1)$ fractional 
quantum Hall effect (FQHE) 
state is mapped on to the $\nu'=p$ integer QHE state of 
composite fermions.
The spin transitions at $\nu=2/3$ and 
2/5\cite{jain2} correspond to the spin transition at $\nu=2$, 
where the Zeeman splitting corresponds to
the effective Landau level separation, and 
the energy levels of the minority spin state in the 
lowest Landau level and the majority spin state 
in the second lowest Landau level coincide.

Extensive experimental\cite{exp1,exp1.1,exp2.1,exp2,exp3,exp4,
exp5,exp7,exp8}
 and theoretical \cite{thry1,thry2,thry3,karel} 
studies have been made on this transition. 
Nevertheless, there is no clear theoretical consensus on this
issue. This is due to the difficulties of
studies in this system.  A number of states 
possibly compete in energy, and large enough systems are needed to 
see non-uniform structures in the partially polarized states.
Here we use the DMRG method\cite{Shibata1},
and study the spin transition and the spin structures in large system 
to clarify the nature of the spin transition at $\nu=2/3$.

\begin{figure}[t]
\begin{center}
\includegraphics[width=0.55\textwidth]{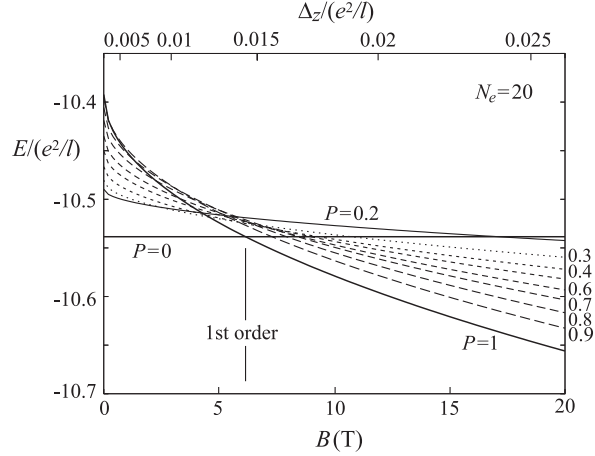}
\caption{\label{spin_e}
Lowest energies for fixed polarization ratio $P$ as a function 
of magnetic field $B$ at filling factor $\nu=2/3$ in units of
 $e^2/(\epsilon \ell)$. 
The total number of electron is 20. The aspect ratio is fixed at 2.0.
The $g$-factor is 0.44.
}
\label{figure1}
\end{center}
\end{figure}

\begin{figure}[t]
\begin{center}
\includegraphics[width=0.48\textwidth]{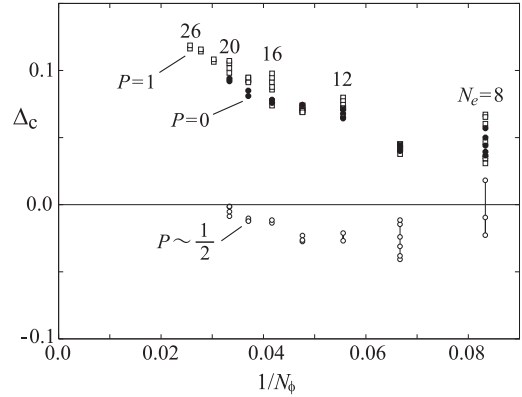}
\caption{\label{spin_gap}
Charge gap of $\nu=2/3$ spin polarized states ($\Box$), unpolarized
 states ($\bullet$), and partially polarized states ($\circ$)
for various $N_e$ and aspect ratios $L_x/L_y$. $\Delta_c$ is
in units of $e^2/(\epsilon \ell)$. 
}
\label{figure2}
\end{center}
\end{figure}

We first calculate the energy at various polarization 
$P$ as a function of the Zeeman splitting, $\Delta_z=g\mu B$. 
The obtained results are shown in Fig.~\ref{spin_e}.
In the absence of the Zeeman splitting,
the unpolarized state ($P=0$) is the lowest.
The energy of the polarized state ($P>0$) monotonically increases 
as $P$ increases. 
With the increase in Zeeman splitting $\Delta_z$, however,  
the energy of polarized state decreases and 
the fully polarized state ($P=1$) becomes the lowest.
Figure \ref{spin_e} shows that the transition from the unpolarized state 
to the fully polarized state occurs at $B\simeq 6$T which is roughly
consistent to the earlier work done in a spherical geometry.  
\cite{chacraborty}
In the present calculation on a torus,
all partially polarized
states ($0< P < 1$) are higher in energy than the ground states ($P=0$
or 1). 
This feature is independent of the size of the system and 
the aspect ratio $L_x/L_y$, and
 indicates phase separations of $P=0$ and 
$P=1$ in partially polarized states.

The unpolarized state of $P=0$ and the fully polarized 
state of $P=1$ are both quantum Hall states 
with finite charge excitation gap, which is
 defined by 
\begin{equation}
 \Delta_c(P)=E(N_{\phi}+1,P)+E(N_{\phi}-1,P)-2E(N_{\phi},P), 
\end{equation}
where $N_{\phi}$ is the number of one-particle states
in the lowest Landau level.
The filling factor $\nu$ is then given by $N_e/N_{\phi}$.
The charge gap $\Delta_c$ for various $N_{\phi}$ 
and aspect ratios of the unit cell is presented in Fig.~\ref{spin_gap}.
In this figure, the gap $\Delta_c$ seems to vanish 
for partially polarized state $P\sim 1/2$
in the limit of $N_e\rightarrow\infty$.
This result clearly indicates that partially polarized 
state with 
$P\sim 1/2$ is a compressible state in contrast to the
incompressible states at $P=0$ and $1$, where
$\Delta_c$ remains to be finite in the limit of
$N_e\rightarrow\infty$.

\begin{figure}[t]
\begin{center}
\includegraphics[width=0.55\textwidth]{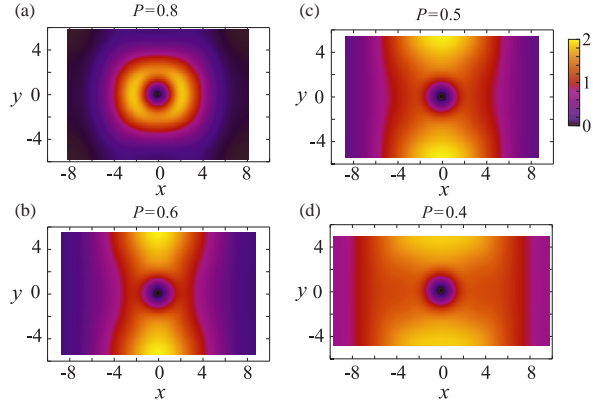}
\caption{\label{spin_dom}
Pair correlation functions for minority spins 
$g_{\downarrow\downarrow}$ at $\nu=2/3$
for several polarization ratios 
(a) $P=0.8$, (b) $P=0.6$, (c) $P=0.5$, and (d) $P=0.4$.
}
\label{figure3}
\end{center}
\end{figure}

To study the spin structure in the partially polarized states, 
we next calculate the pair-correlation function defined by
\begin{equation}
 g_{\sigma\sigma}({\mib r})=\frac{L_xL_y}{N_{\sigma}(N_{\sigma}-1)}
\langle \Psi |\sum_{nm}\delta({\mib r}+{\mib R}_{\sigma,n}-{\mib R}_{\sigma,m})
|\Psi\rangle ,
\end{equation}
where $\sigma=\pm1/2$ is the spin index and $N_{\sigma}$ is the number
of electrons with spin $\sigma$.
The spin structures in partially spin polarized states 
are clearly shown in 
the pair correlation function between minority spins. 
Namely, if unpolarized regions are formed
in the partially polarized states, then 
electrons with minority spins 
are concentrated in the unpolarized regions.
This concentration of the minority spins is 
shown in Fig.~\ref{spin_dom}, which shows
$g_{\downarrow\downarrow}(x,y)$ for partially polarized states
at (a) $P=0.8$, (b) $P=0.6$, (c) $P=0.5$, and (d) $P=0.4$. 
When $P$ is close to $1$, for example $P=0.8$ shown
in Fig.~\ref{spin_dom}(a), a pair of minority spins is found
only near the origin. 
As the polarization ratio $P$ decreases, minority 
spins make a domain around the origin, and two domain walls 
along the $y$-direction are formed.
These domain walls move along $x$-direction
and the domain of minority spin finally 
covers entire unit cell in the limit of $P=0$.
This change in the size of the domain is consistent with the expectation
that the domain in Fig.~\ref{spin_dom} 
corresponds to the unpolarized spin singlet region
where the density of up-spin electrons
and the down-spin electrons are the same.

\begin{figure}[t]
\begin{center}
\includegraphics[width=0.55\textwidth]{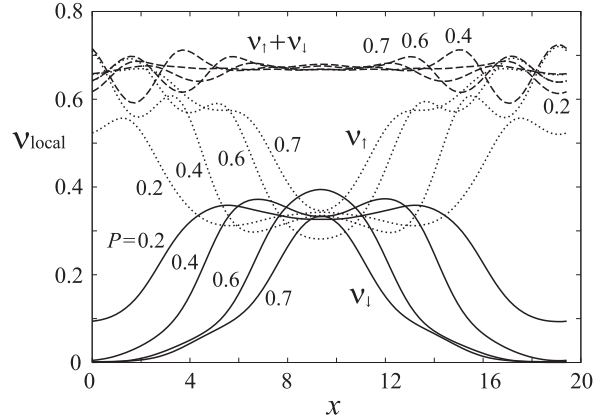}
\caption{\label{domain_4}
Local densities of up spin, and down spin electrons
for various polarization ratios $P$ at $\nu=2/3$. 
The number of electrons is 20.
}
\label{figure4}
\end{center}
\end{figure}

To confirm the separation of the unpolarized and polarized spin regions, 
we next consider the local electron density 
of up-spin electrons $\nu_{\uparrow}(x)$ and down-spin electrons 
$\nu_{\downarrow}(x)$. 
Figure \ref{domain_4} shows  $\nu_{\uparrow}(x)$ and $\nu_{\downarrow}(x)$
for partially polarized states with $P=0.2,\ 0.4,\ 0.6$ and $0.7$. 
Here $\nu_{\uparrow}(x)$ and $\nu_{\downarrow}(x)$ are
scaled to be the local filling factor of the 
lowest LL. Thus, the total local electron density 
$\nu_{\uparrow}(x)+\nu_{\downarrow}(x)$ is almost $2/3$.
In this figure the separation to two regions is clearly seen;
the unpolarized spin region around $L_x/2$, where 
both $\nu_{\uparrow}$ and $\nu_{\downarrow}$ are close to 1/3,
and the fully polarized spin region around $x\sim 0$ or equivalently
 $x\sim L_x$, where $\nu_{\uparrow}$ is almost 2/3 while
 $\nu_{\downarrow}$ is close to 0.
These results confirm the separation of the unpolarized and 
polarized spin regions as expected from the 
pair correlation functions shown in Fig.~\ref{domain_4}.

The polarized and unpolarized spin regions are separated  
by the domain walls whose width is about $4\ell$. This means that the 
phase separation is realized only for systems whose
size of the unit cell $L_x , (L_y) $ is larger than twice the width of
domain wall; $L_x,(L_y) > 8\ell$. 
Indeed, exact diagonalization studies up to $N_e=8$ electrons have 
never found the phase separation at $\nu=2/3$.\cite{karel} 
We have found the phase separation only for large systems with 
$N_e > 12$. 
We note that above behavior is generic over the aspect ratio.
In an ideal system, the two states separate into two regions even when the
system size is infinitely large. In experimental situations, however, 
multi-domain structures are realized due to the inhomogeneity and
coupling with randomly distributed nuclear spins.

The DMRG study on the ground state energy for various polarization $P$ 
shows that the ground state at $\nu=2/3$
evolves discontinuously from the unpolarized $P=0$ 
state to the fully polarized $P=1$ state as the Zeeman splitting increases.
In partially polarized states $0<P<1$, the electronic system separates 
spontaneously into two states; the $P=0$ and the $P=1$ states. 
These two states are separated by the domain wall of width $4\ell$.
Since the energy of the domain wall is positive,  
the partially polarized states always have higher energy than
that of $P=1$ or $P=0$ states. 
We think this is the reason of the direct first order transition 
from  $P=0$ to $P=1$ state in the ground state.

It is useful to compare our result with the spin transition 
at $\nu=2$ which occurs when minority spin states in the lowest LL 
and majority spin states in the second lowest LL cross 
by varying the ratio of the Zeeman and Coulomb energy.
The ground state at $\nu=2$ is thus a fully polarized state 
or a spin singlet state. 
In analogous to the case of $\nu=2/3$ the transition between them is 
first order\cite{tomas}, 
and spin domain walls have been found in high energy states.\cite{nomura}
This analogy can be expected, because the $\nu=2$ states and the $\nu=2/3$ 
states are connected in the composite fermion theory\cite{jain,jain2}, 
although the effective interaction between composite fermions is different 
from that for electrons.

\section{Bilayer system}

The properties of quantum Hall systems sensitively depend on the 
magnetic field, and various types of ground states including
incompressible liquids\cite{Laughlin}, compressible 
liquids\cite{Jain,Halp}, spin singlet liquid,
CDW states called stripes, bubbles, and
Wigner crystal are realized depending on the filling $\nu$
of Landau levels.
In bilayer quantum Hall systems, additional length scale of 
the layer 
distance $d$, and the degrees of freedom of layers make the 
ground state much more diverse and interesting.\cite{EM}

Excitonic phase, namely Haplerin's $\Psi_{1,1,1}$ state, 
is one of the ground states realized in
bilayer quantum Hall systems at total filling $\nu=1$ at small 
layer separation $d$, where electrons and holes in different layers
are bound with each other due to strong
interlayer Coulomb interaction.
This excitonic state has recently attracted much attention
because a dramatic enhancement of zero bias tunneling conductance 
between the two layers\cite{ZTNC}, and the vanishing of the 
Hall counterflow resistance are observed\cite{CFH1,CFH2}.
As the layer separation $d$ is increased, the excitonic phase vanishes, 
and at large enough separation, composite-fermion Fermi-liquid 
state is realized in each layer. 

Several scenarios have been proposed for the transition
of the ground state as the layer separation 
increases\cite{HFT,PCF,Mac,Kim,SH,NY,SRM}.
However how the excitonic state develops into 
independent Fermi-liquid state has not been fully understood. 
In this section we investigate the ground state of $\nu=1$ bilayer 
quantum Hall systems by using the DMRG method\cite{Shibata1}. 
We calculate energy gap, two-particle correlation function $g(r)$ 
and excitonic correlation function for 
various values of layer separation $d$, and show the 
evolution of the ground state with increasing $d$.

The Hamiltonian of the bilayer quantum Hall systems is written by
\begin{eqnarray}
H &=& \sum_{i<j} \sum_{\mib q} V(q)\ {\rm e}^{-q^2\ell^2/2} 
{\rm e}^{{\rm i}{\mib q} \cdot ({\mib R}_{1,i}-{\mib R}_{1,j})} \nonumber \\
&&+ \sum_{i<j} \sum_{\mib q} V(q)\ {\rm e}^{-q^2\ell^2/2} 
{\rm e}^{{\rm i}{\mib q} \cdot ({\mib R}_{2,i}-{\mib R}_{2,j})} \nonumber \\
&&+ \sum_{i,j} \sum_{\mib q} V(q)\ {\rm e}^{-qd}e^{-q^2\ell^2/2} 
{\rm e}^{{\rm i}{\mib q} \cdot ({\mib R}_{1,i}-{\mib R}_{2,j})},
\label{Coulomb}
\end{eqnarray}
where ${\mib R}_{1,i}$ are the two-dimensional guiding center 
coordinates of the $i$th electron in the layer-1 and 
${\mib R}_{2,i}$ are that in 
the layer-2. The guiding center coordinates
 satisfy the commutation relation,
$[{R}_{j}^x,{R}_{k}^y]={\rm i}\ell^2\delta_{jk}$.
$V(q) =2\pi e^2/(\epsilon q)$ is the Fourier transform of the
Coulomb interaction and 
the wave function is projected on to the lowest Landau level.
We consider uniform positive background charge to cancel the
component at $q=0$.
We will assume zero interlayer tunneling and fully spin polarized 
ground state.

\begin{figure}[t]
\begin{center}
\epsfxsize=80mm \epsffile{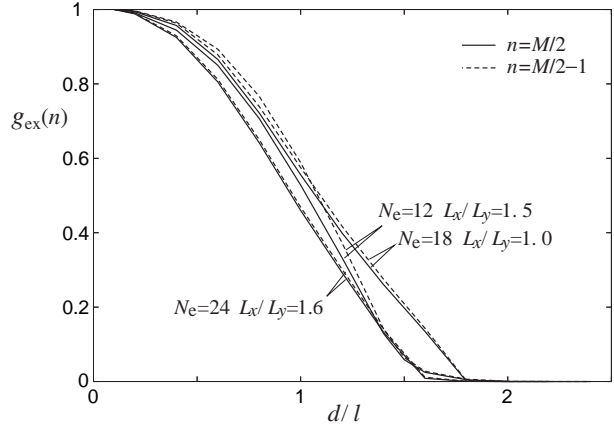}
\caption{\label{bi_cor}
The exciton correlation of bilayer quantum Hall systems at
$\nu=1$. The solid line represents $g_{\rm ex}(M/2)$.
The dashed line represents $g_{\rm ex}(M/2-1)$.
}
\end{center}
\end{figure}

In the limit of $d=0$, electrons in different layers
can not occupy the same position 
because of the strong interlayer Coulomb repulsion. 
The strong interlayer repulsion makes electron-hole 
pairs, which is called excitons whose 
degrees of freedom are represented by interlayer dipoles or
pseudo-spins at total filling $\nu=1$. 
The Coulomb exchange interaction aligns the interlayer 
dipoles (the pseudo-spins) leading to the macroscopic
coherence of the excitons and Haplerin's $\Psi_{1,1,1}$ 
state is realized.

To confirm the coherence of the excitons,
we calculate the exciton correlation defined by 
\begin{equation}
g_{\rm ex}(n) \equiv
\frac{2M-1}{N_1N_2}
\langle \Psi | c^\dagger_{1,n} c_{2,n}  
c^\dagger_{2,0} c_{1,0}  |\Psi \rangle,
\end{equation}
where $|\Psi\rangle$ is the ground state and $c^\dagger_{1,n}$ 
($c^\dagger_{2,n}$) is the creation operator of
the electrons in the $n$th one-particle state defined by
\begin{equation}
\phi_n({\mib r})=\frac{1}{\sqrt{L_y\pi^{1/2}\ell}}
\exp\left\{i k_y y - \frac{(x-X_n)^2}{2\ell^2}\right\}
\end{equation}
in the layer-1 (layer-2). 
$X_n=nL_x/M=k_y\ell^2$ is the $x$-component of the guiding 
center coordinates and 
$L_x$ is the length of the unit cell in the $x$ direction. 
$M$ is the number of one-particle state in each layer.
$N_1$ and $N_2$ are the number of 
electrons in the layer-1 and layer-2, respectively, and
we impose the symmetric condition of $N_1=N_2$.
Since $g_{\rm ex}(n)$ represents the correlations between the 
two excitons at 
$X=0$ and $X=X_n$, $\lim_{n\rightarrow \infty} g_{\rm ex}(n)\ne 0$ indicates 
existence of macroscopic coherence of excitons. 

As is shown in Fig.~\ref{bi_cor}, $g_{\rm ex}(n)$ tends to 1 as $d \to 0$,
that confirms the macroscopic coherence of excitons at $d=0$.
Indeed, Haplerin's $\Psi_{1,1,1}$ state has the macroscopic
coherence of excitons and $g_{\rm ex}(n)=1$ independent of $n$. 
In this figure we have shown $g_{\rm ex}(M/2)$ instead of 
$\lim_{n\rightarrow \infty} g_{\rm ex}(n)$, 
because the largest distance between the two excitons is 
$L_x/2$ in the finite unit cell of $L_x\times L_y$ under the
periodic boundary conditions. 
In order to check the size effect, we also plot $g_{\rm ex}(M/2-1)$
with the dashed line. Since the difference between 
$g_{\rm ex}(M/2)$ and $g_{\rm ex}(M/2-1)$ is small, we expect 
$g_{\rm ex}(M/2)$ well represents the macroscopic coherence
in the limit of $N\rightarrow \infty$.
With increasing $d/\ell$, the excitonic correlation decreases monotonically 
and finally falls down to negligible value at $d/\ell \sim 1.6$. 

\begin{figure}[t]
\begin{center}
\epsfxsize=80mm \epsffile{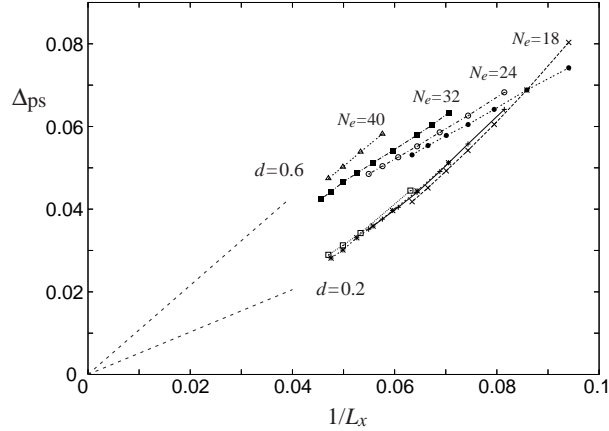}
\caption{\label{bi_sgap}
The  pseudo-spin excitation gap $\Delta_{ps}$ 
of bilayer quantum Hall systems at the total filling 
factor $\nu=1$. The dashed lines are guide for the eye.
}
\end{center}
\end{figure}

The presence of macroscopic coherence of excitons shown in the 
Fig.~\ref{bi_cor} 
means the existence of ferromagnetic order of the 
interlayer dipoles (the pseudo-spins). Since interaction between 
the interlayer dipoles has SU(2) symmetry at $d=0$, 
corrective gapless excitations called pseudo-spin waves 
are expected. Even in the case of finite layer distance $d$,
the Hamiltonian has continuous XY symmetry, and 
gapless pseudo-spin wave excitations are still expected.
This is confirmed by the size dependence of the pseudo-spin
excitation gap shown in Fig.~\ref{bi_sgap}, where
the pseudo-spin excitation gap $\Delta_{ps}=E(N_1+1,N_2-1,M)-E(N_1,N_2,M)$ 
in finite system decreases as a function of $1/L_x$.

\begin{figure}[t]
\begin{center}
\epsfxsize=80mm \epsffile{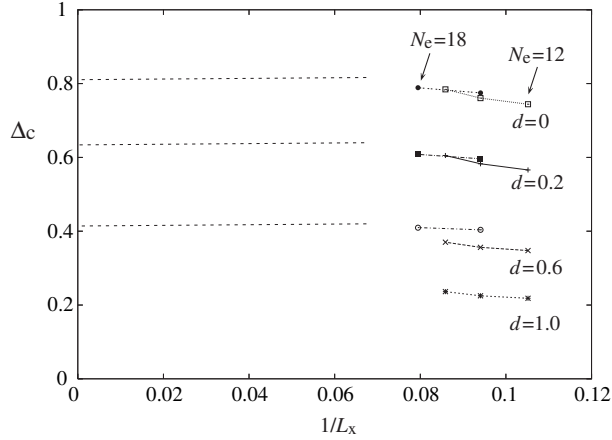}
\caption{\label{bi_cgap}
The charge excitation gap $\Delta_{c}$ 
of bilayer quantum Hall systems at the total filling 
factor $\nu=1$. The dashed lines are guide for the eye.
}
\end{center}
\end{figure}

In contrast to the pseudo-spin excitation gap $\Delta_{ps}$, 
the charge excitation gap $\Delta_c$ defined by 
$\Delta_{c}=E(N_1,N_2,M-1)+E(N_1,N_2,M+1)-2E(N_1,N_2,M)$
seems to be finite even in the limit of $L_x \rightarrow \infty$
for small $d$ as shown in Fig.~\ref{bi_cgap}.
The charge excitation brakes at least one electron-hole pair
and it needs energy of order $V_0^{(1,2)}$ which is the 
pseudopotential between the electrons in different layers
whose relative angular momentum is $0$. 
This pseudopotential decreases with the increase in the layer 
distance $d$, and thus the charge gap decreases with 
 the increase in $d$.   

\begin{figure}[t]
\begin{center}
\epsfxsize=80mm \epsffile{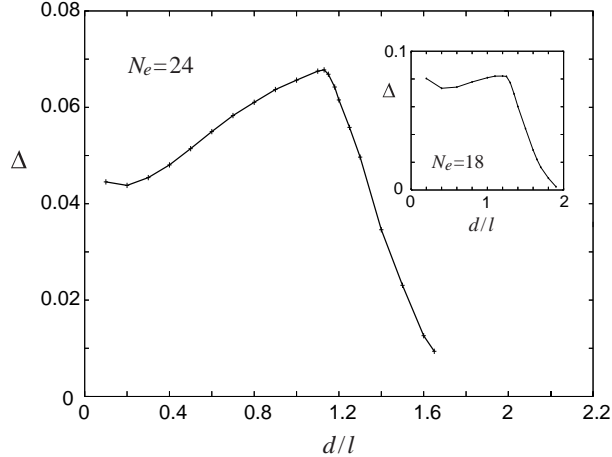}
\caption{\label{bi_gap}
The lowest excitation gap $\Delta$
of bilayer quantum Hall systems at the total filling 
factor $\nu=1$. $N_e=24$ and $L_x/L_y=1.6$.
The inset shows the result for $N_e=18$ and $L_x/L_y=1.0$.
}
\end{center}
\end{figure}

We next see the lowest excitation gap in a fixed size of system. 
Figure~\ref{bi_gap} shows the result for $N_e=24$. 
The aspect ratio $L_x/L_y=1.8$ is chosen from the minimum of 
the ground state energy with respect to $L_x/L_y$ around
$d/\ell=1.8$, where minimum structure appears in the ground state
energy.

We can see clear excitation gap of finite system for $d/\ell<1.2$,
where excitonic ground state
is expected both theoretically and experimentally
\cite{ZTNC,CFH1,CFH2,HFT,PCF,Mac,Kim,SH,NY,SRM}.
The excitation gap rapidly decreases with increasing  $d/\ell$
from $1.2$, and it becomes very small for $d/\ell>1.6$. 
This behavior is consistent with experiments.\cite{Wies}
Although the excitation gap for $d/\ell> 1.7$ is not presented 
in the figure for $N_e=24$ because of the difficulty of the calculation
of excited states in large system, we do not find any sign 
of level crossing in 
the ground state up to $d/\ell\sim 4$, where two layers are 
almost independent.
These results suggest that the excitonic state at small $d/\ell$
continuously crossovers to compressible state at large $d/\ell$,
that is consistent with the behaviors of exciton correlations
$g_{\rm ex}(M/2)$ in Fig.~\ref{bi_cor}, which shows 
$g_{\rm ex}(M/2)$ 
continuously approaches zero around $d/\ell \sim 1.6$.
In the present calculation it is difficult to conclude whether 
the gap closes at finite $d/\ell\sim 1.6$ in the thermodynamic limit 
or excitonic state survives with exponentially small finite gap
even for large $d/\ell>1.6$.
We have calculated the excitation gap
in different size of systems and aspect ratios, 
and obtained similar results as shown in the inset of Fig.~\ref{bi_gap}.

Concerning the first excited state, however, Fig.~\ref{bi_gap}. 
shows a level crossing at $d/\ell\sim 1.2$,
where we can see sudden decrease in the excitation gap.
We expect that the lowest excitation at  $d/\ell < 1.2$ is 
the pseudo-spin excitation whose energy gap decreases with the increase
in the size of system and tends to zero 
in the limit of large system.
On the other hand, the lowest excitation at $d/\ell > 1.2$ shown 
in Fig.~\ref{bi_gap}
is expected to be the excitation to the roton minimum 
which corresponds to the bound state of 
quasiparticle and quasihole excitatins, whose energy
increases with the decrease in $d/\ell$.
This change in the character of the low energy excitations 
at $d/\ell \sim 1.2$ will be confirmed 
in a clear change in correlation functions in the excited 
state as shown later. We note that the position of the 
level crossing in the first excited state itself depends on 
the size of system because the pseudo-spin excitation gap 
decreases with the increase in the system size. However, 
the change in the character of the low energy excitations 
of finite systems is expected to remain even in the limit 
of large system, since the spectrum weight of pseudo-spin 
waves transfers to high energy with the increase in $d$.

\begin{figure}[t]
\begin{center}
\epsfxsize=80mm \epsffile{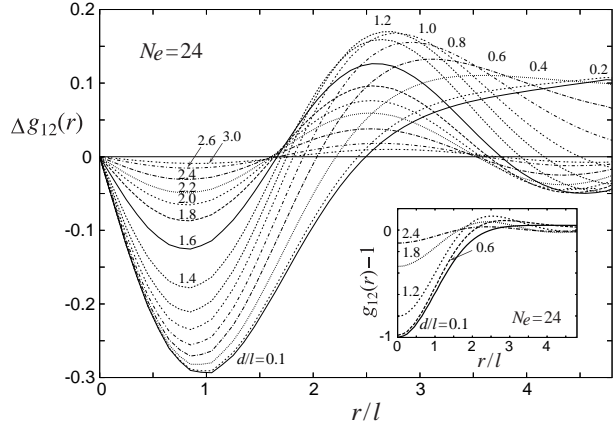}
\caption{\label{bi_inter} 
The inter-layer pair correlation function of electrons
in the ground state of bilayer quantum Hall systems at $\nu=1$. 
$N_e=24$ and $L_y/L_x=1.6$.
}
\end{center}
\end{figure}

We next calculate pair correlation functions of the electrons to
see detailed evolution of the ground-state wave function. 
The interlayer pair correlation functions are defined by 
\begin{eqnarray}
g_{12}({\mib r}) &\equiv& \frac{L_x L_y}{N_1N_2}\langle 
\Psi | \sum_{n\ m} \delta({\mib r}+{\mib R}_{1,n}
-{\mib R}_{2,m})|\Psi 
\rangle,
\end{eqnarray}
where $|\Psi\rangle$ is the ground state.
We present $\Delta g_{12}(r)$ in Fig.~\ref{bi_inter}, which is defined by 
\begin{eqnarray}
\Delta g_{12}(r) &=& \int  (g_{12}({\mib r'})-1)\delta(|{\mib r'}|-r) 
\ {\rm d}{\mib r'}, 
\end{eqnarray}
where ${\mib r'}$ is the two-dimensional position vector in each layer.
$\Delta g_{12}(r)$ represents the difference from the uniform correlation 
of independent electrons.

At $d/\ell=0$ we find clear negative $\Delta g_{12}(r)$ around 
$r/\ell =1$, which 
is a characteristic feature of the excitonic state made by
the binding of electrons and holes between the two layers.
The binding of one hole means the exclusion of one electron
caused by the strong interlayer Coulomb repulsion.
The increase in the layer separation weakens 
Coulomb repulsion between the two layers and 
reduces $|\Delta g_{12}(r)|$ around $r/\ell =1$.

\begin{figure}[t]
\begin{center}
\epsfxsize=80mm \epsffile{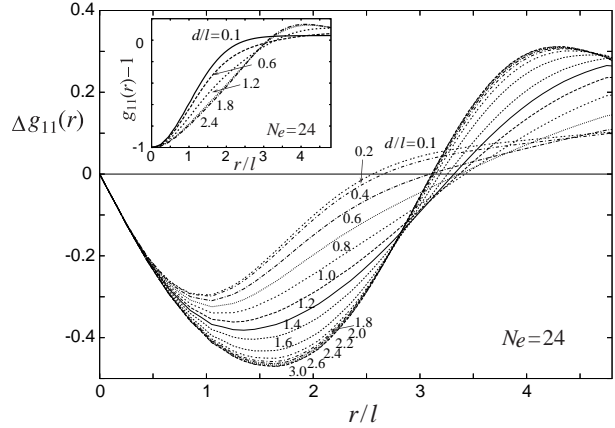}
\caption{\label{bi_intra}
The intra-layer pair correlation function of electrons
in the ground state of bilayer quantum Hall systems at $\nu=1$. 
$N_e=24$ and $L_y/L_x=1.6$. 
}
\end{center}
\end{figure}

The decrease in the interlayer correlation $|\Delta g_{12}(r)|$ 
opens space to enlarge correlation hole in the same layer
and reduce the Coulomb energy between the electrons within the layer.
This is shown in Fig.~\ref{bi_intra}, which shows the pair correlation 
functions of the electrons in the same layer defined by
\begin{eqnarray}
g_{11}({\mib r}) &\equiv& \frac{L_x L_y}{N_1(N_1-1)}\langle 
\Psi | \sum_{n m} \delta({\mib r}+{\mib R}_{1,n}-{\mib R}_{1,m})|\Psi
\rangle, 
\end{eqnarray}
\begin{eqnarray}
\Delta g_{11}(r) &=& \int  
(g_{11}({\mib r'})-1)\delta(|{\mib r'}|-r) \ {\rm d}{\mib r'} .
\end{eqnarray}
The obtained results indeed show
that the correlation hole in the same layer
around $r/\ell \sim 1$ is enhanced with the increase in 
$d/\ell$ contrary to the decrease in size of
interlayer correlation hole in Fig.~\ref{bi_inter}. 
The correlation hole in the same layer monotonically increases in size 
up to $d/\ell=1.8$, and then it becomes almost constant. 
The correlation function $g_{11}(r)$ for $d/\ell>1.8$
is almost the same to that of $\nu=1/2$ monolayer quantum Hall 
systems realized in the limit of $d/\ell=\infty$. 
This is consistent with the almost vanishing 
excitation gap and exciton correlation at $d/\ell>1.8$ 
shown in Figs.~\ref{bi_gap} and \ref{bi_cor}. 

Figure \ref{bi_intra} also shows that the growing of
the correlation hole around $r/\ell\sim 1.5$ is 
accompanied with the increase in 
$\Delta g_{11}(r)$ around $r/\ell\sim 4$. 
The distance $4\ell$ is comparable to the approximate 
mean distance between the electrons $3.54\ell$ 
estimated from $(L_xL_y/N_1)^{1/2}= (2\pi L/N_1)^{1/2} \ell$.
This means the electrons in the same layer tend to 
keep distance of about $4\ell$ from other electrons 
with the large correlation hole around $r/\ell\sim 1.5$ 
for $d / \ell  \stackrel{>}{_\sim} 1$. 
This is consistent with the formation of composite fermions 
at $d=\infty$, 
where two magnetic flux quanta are attached to each electron, 
which is equivalent to enhance the correlation hole 
around each electron in the same layer to keep distance
from other electrons.

The large correlation hole in $g_{11}(r)$ 
attracts electrons in the other layer as shown in Fig.~\ref{bi_inter}, 
where we find a clear peak in $\Delta g_{12}(r)$ at $r/\ell \sim 3$.
This peak at $r/\ell \sim 3$ is comparable to the neighboring
correlation hole at $r/\ell \sim 1$, which suggests that the 
electrons excluded from the origin by strong interlayer Coulomb 
repulsion are trapped by the correlation hole in $g_{11}(r)$ 
within $r/\ell\sim 4$.
Since the intra-layer correlation $g_{11}$ for $d/\ell>1.6$
is almost the same to that of composite-fermion liquid state, 
$\Delta g_{12}(r)$ represents the correlation of composite 
fermions between the layers. 
The almost same amplitude of $\Delta g_{12}(r)$ at 
$r/\ell \sim 1$ and $3$ for $d/\ell>1.6$ actually shows that
the electrons in the other layer bind holes to 
form composite fermions. 

With decreasing $d/\ell$ from infinity, the 
correlations of composite fermions in 
different layers monotonically increases down to 
$d/\ell\sim 1.2$ as shown in the enhance of 
$|\Delta g_{12}(r)|$ at $r/\ell \sim 1$ and $3$. 
But further decrease in $d/\ell$ broadens 
the peak at $r/\ell \sim 3$ in $\Delta g_{12}(r)$ 
with the decrease in the correlation hole in $\Delta g_{11}(r)$, 
and the peak at $r/\ell \sim 3$ in $\Delta g_{12}(r)$ 
finally disappears.
This change in the correlation function shows how the 
composite-fermion liquid state evolves into excitonic state:
The large correlation hole in the same layer, which is a 
characteristic feature of the composite fermions, 
is transfered into the other layer to form excitonic state. 
The correlation functions in Figs.~\ref{bi_inter} and \ref{bi_intra} are
continuously modified with the decrease in $d/\ell$ from 
$\infty$ to $0$, which supports continuous 
transition from the compressible liquid state to the 
excitonic state.
Fig.~\ref{bi_inter} also shows that the peak in $\Delta g_{12}(r)$ at 
$r/\ell \sim 3$ made by the binding of an electron to 
the hole around the origin gradually disappears with
decreasing $d/\ell$ from 1.2.
This means the gradual break down of the concept of 
composite fermions.

\begin{figure}[t]
\begin{center}
\epsfxsize=80mm \epsffile{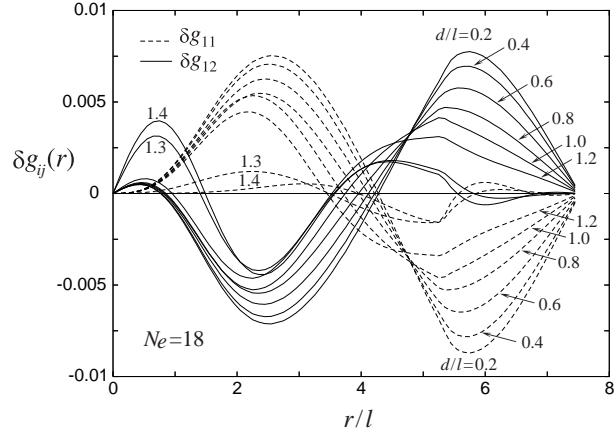}
\caption{\label{bi_exc}
The change in correlation function through the excitation
from the ground state to the first excited state.
$\nu=1$ and $N_e=18$ with $L_y/L_x=1.0$.
}
\end{center}
\end{figure}

The break down of the composite fermions around $d/\ell \sim 1.2$
affects the character of the lowest excitations, 
which is clearly shown in the 
level crossing of the excited state at $d/\ell\sim 1.2$.
The change in the character of excitation
is confirmed by the correlation functions in the
excited state. Figure \ref{bi_exc} shows the difference 
in the pair correlation 
functions $g_{ij}(r)$ between the ground state
and first excited state defined by
\begin{eqnarray}
\delta g_{ij}(r) &=& \int  
(g_{ij}^{E}({\mib r'})-g_{ij}^{G}({\mib r'}))
\delta(|{\mib r'}|-r) \ {\rm d}{\mib r'} ,
\end{eqnarray}
where $g_{ij}^{G}({\mib r})$ and $g_{ij}^{E}({\mib r})$ are
the pair correlation functions 
in the ground state and the first excited state, respectively.
$\delta g_{ij}(r)$ in Fig.~\ref{bi_exc} show that there is a discontinuous 
transition between $d/\ell=1.2$ and $1.3$, which supports
the level crossing in the first excited state. 

Below $d/\ell\sim 1.2$, $\delta g(r)$ have large amplitude
at $r/\ell\sim 2$ and $6$, which shows electrons are 
transfered between the inside of $r/\ell\sim 4$ and its outside.
Small singularity at $r/\ell \sim 5.5$ is due to finite
size effects of square unit cell.
Above $d/\ell\sim 1.2$, only $\delta g_{12}(r)$ have 
large amplitude at $r/\ell\sim 1$ and 2, which shows 
the electrons within $r/\ell\sim 4$ in different layers 
are responsible for the lowest excitation.
This result suggests that the low energy excitations are 
made by composite fermions in different layers for $d/\ell > 1.2$.

\section{Conclusions}

In this paper we have reviewed the ground state and 
low energy excitations of the quantum Hall systems studied 
by the DMRG method. 
We have applied the DMRG method to two dimensional quantum 
systems in magnetic field by using a mapping on to an 
effective one-dimensional lattice model.
Since the Coulomb interaction between the electrons is 
long-range, all the electrons in the system interact 
with each other. This fact seems to severely reduce 
the accuracy of the DMRG calculations. However, in the 
magnetic field, one-particle wave functions are
localized within the 
magnetic length $\ell$, and the overlap of the one-particle
wave functions exponentially decreases
with increasing the distance between the two electrons.
This means the quantum fluctuations are restricted to
short-range and the effective Hamiltonian is suited for the 
DMRG scheme. This is the reason why relatively small
number of keeping states is enough for quantum Hall systems
compared with usual two dimensional systems.

In quantum Hall systems, filling $\nu$ of Landau levels 
is determined by $\nu=N_e/N_\phi$, where $N_\phi$ is the number
of flux quanta and related to the magnetic field
as $N_\phi=(e/h)L_xL_y B $.
Thus so many types of the ground state
are realized only by changing the uniform magnetic field $B$.
Since the ground state of free electrons in 
partially filled Landau level has macroscopic degeneracy, 
Coulomb interaction drastically changes the 
wave function. The character of the ground state 
is sensitive to the Landau level index $N$ and the filling $\nu$,
which modify the effective interaction and the mean distance 
between the electrons.
This is the source of many interesting low temperature 
properties of quantum Hall systems and their inherent difficulties.

\section*{Acknowledgments}
The author would like to thank Prof. Yoshioka Daijiro
and Dr. Kentaro Nomura for valuable discussions.
This work is supported by 
Grant-in-Aid No. 18684012 from MEXT, Japan.

%

\end{document}